\documentclass[%
 aip,
 jmp,%
 amsmath,amssymb,
 reprint,%
]{revtex4-2}

\usepackage{graphicx}
\usepackage{dcolumn}
\usepackage{bm}

\begin{document}


\title[Non-equilibrium turbulent transport in convective plumes]{Non-equilibrium turbulent transport in convective plumes obtained from closure theory}

\author{N. Yokoi}
 \altaffiliation[]{Visiting researcher at the Nordic Institute for Theoretical Physics (NORDITA).}
\affiliation{ 
Institute of Industrial Science (IIS), University of Tokyo, 4-6-1, Komaba, Meguro, Tokyo 153-8505, Japan
}%

\date{\today}

\begin{abstract}
Non-equilibrium property of turbulence modifies characteristics of turbulent transport. With the aid of response-function formalism, such non-equilibrium effects in turbulent transport can be represented by the temporal variation of the turbulent energy ($K$) and its dissipation rate ($\varepsilon$) along the mean stream through the advective derivatives of $K$ and $\varepsilon$. Applications of this effect to the turbulent convection with plumes are considered for the first time in this work. The non-equilibrium transport effects associated with plumes are addressed in two aspects. Firstly, the effect associated with a single plume is evaluated using data measured in the recent plume/jet experiments. The second argument is developed for the collective turbulent transport associated with multiple plumes mimicking the stellar convection zone. In this second case, for the purpose of capturing the plume motions into the advective derivatives, use has to be made of the time--space double averaging procedure, where the turbulent fluctuations are divided into the coherent or dispersion component (which represents plume motions) and incoherent or random component. With the aid of the transport equations of the coherent velocity stress and the incoherent counterpart, the interaction between the dispersion and random fluctuations are also discussed in the context of convective turbulent flows with plumes. It is shown from these analyses that the non-equilibrium effect associated with plume motions is of a great deal of relevance in the convective turbulence modelling.
\end{abstract}

\maketitle
\def\NOTE#1{{\textcolor{blue}{#1}}}




\setcounter{section}{0} 
\section{Introduction \label{sec:intro}}
Turbulent flows encountered in astro/geophysical phenomena as well as laboratory plasma are ubiquitously at huge Reynolds number. Since direct numerical simulations (DNSs) of such flows are just impossible in the foreseeable future, turbulence or subgrid-scale (SGS) modeling approach provides a powerful tool for analyzing turbulent flows in these phenomena. 

In the conventional turbulence modeling, simple expressions of turbulent fluxes such as the eddy-viscosity and eddy-diffusivity representations are very useful and extensively utilized.\citep{lau1972} For instance, the turbulent momentum flux, defined by the Reynolds stress, is modeled by the turbulent transport coefficients, eddy viscosity, coupled with the mean velocity strain. The eddy viscosity is expressed in terms of the quantities that characterize dynamics and statistical properties of the turbulence. In the mixing-length formulation, the eddy viscosity $\nu_{\rm{T}}$ is expressed by the characteristic turbulent velocity $v$ and length scale $\ell$ as $\nu_{\rm{T}} = \nu_{\rm{T}}(v, \ell)$. In the standard $K-\varepsilon$ model, $\nu_{\rm{T}}$ is modeled in terms of the turbulent kinetic energy $K$ and its dissipation rate $\varepsilon$ as $\nu_{\rm{T}} = \nu_{\rm{T}}(K,\varepsilon)$. As compared with the mixing-length formulation, the $K-\varepsilon$ is more elaborated since the dynamics of the turbulent fields and their interaction with the mean fields are taken into account through the transport equations of $K$ and $\varepsilon$.\citep{rod2000,yok2020} 

In spite of the notable achievements of simple gradient-diffusion-type models for turbulent fluxes with the eddy-viscosity and eddy-diffusivity representations, there exist several situations where such models would completely fail. One of such deficiencies shows up when the eddy-viscosity model is applied to a situation with cross-flow configuration. This failure is caused by the lack of vorticity or rotation effects in the eddy-viscosity model. An alleviation of this deficiency of the eddy-viscosity model was proposed with the aid of an analytical statistical theory of inhomogeneous turbulence applied to non-reflectional symmetric system with emphasis on the helicity effects.\citep{yok1993,yok2016} Another kind of deficiency shows up in the case of turbulence with non-equilibrium properties. For instance, in homogeneous shear turbulence, the evaluation of the eddy-viscosity systematically fails because of the overestimates of the turbulent energy $K$ and its dissipation rate $\varepsilon$. There, the energy injection increases with time, and the energy dissipation process cannot catch up with the injection at the energy-containing scale. This is caused by the non-equilibrium property of turbulence induced by the energy injection due to flow shear. In order to alleviate this flaw in the framework of the $K-\varepsilon$ model with the eddy-viscosity representation, the non-equilibrium effect has to be incorporated into the eddy-viscosity representation.\citep{yos1993} 

Employing the response function formalism in turbulence closure studies provides a powerful way to construct a self-consistent turbulence theory without resorting to any parameter. There are several classical modern theories of turbulence closure.\citep{kra1959,edw1964,her1965,her1966} Starting with the Liouville equation derived by \citet{edw1964}, \citeauthor{her1965} \citep{her1965,her1966} constructed a series of the self-consistent field (SCF) theories for turbulence closure. In the formulation, the coupled equations of the two-time velocity correlation function and the response function are derived under the requirement that the probability distribution functions satisfy the self-consistency condition. The obtained equations for the correlation and response functions are similar to the those obtained with the direct-interaction approximation (DIA) by \citet{kra1959}. In the combination of the DIA with multiple-scale analysis, we address the convective turbulence with variable density and velocity shear. One important consequence of this multiple-scale DIA analysis lies in the point that it enables us to treat the non-equilibrium effect of turbulence through the Lagrangian derivative of the turbulence fields. If the turbulence properties such as the turbulent energy $K$ and its dissipation rate $\varepsilon$ are spatially developing in the streamwise direction, the non-equilibrium effects are represented by the advective derivatives of $K$ and $\varepsilon$, $({\bf{U}} \cdot \nabla) K$ and $({\bf{U}} \cdot \nabla) \varepsilon$ in the Lagrangian derivatives. In the model for the homogeneous-shear turbulent flow\citep{yos1993}, the non-equilibrium properties of turbulence including the variations of time and length scales of turbulence are taken into account in the expression of the eddy viscosity through the Lagrangian derivative along the mean velocity ${\bf{U}}$ of the turbulent energy and its dissipation rate as $\nu_{\rm{T}} = \nu_{\rm{T}}(K,\varepsilon,DK/Dt,D\varepsilon/Dt)$ where $D/Dt = \partial/\partial t + {\bf{U}} \cdot \nabla$.

There are several types of buoyant streams in nature. Some buoyant flows are driven purely by thermo-buoyancy (plumes) and others are driven by combination of thermobuoyancy and dynamical forcing (thermobuoyant jets, nonisothermal shear layers). In the buoyancy-dominated turbulent flows, the turbulent convection is well represented by large-scale plumes.\citep{spr1997,ras1998,cos2016,bra2016} In order for a turbulence model to treat turbulent convective flows, the turbulence interaction with coherent structures such as large-scale jet and plume has to be properly represented in the model. Because of the presence of jets and plumes, the turbulent fluxes deviate from the usual gradient diffusivity, viscosity, etc.\citep{bra2016} The turbulent transport coefficients should properly reflect such coherent structure effects. Another example in astrophysics can be seen in the supernova explosion studies. In the core collapse supernovae (CCSNe) explosion problem, turbulent effects aid the relaunch of blast shock waves due to the neutrino heating.\citep{bet1985,jan1996,mar2009,mur2011,mab2018,mue2019,cou2020} In order to identify the turbulence effects most relevant to the explosion condition, \citet{mur2011} extensively examined several types of closure models. By comparing the model simulation results with the 2D simulation counterparts, they found that the standard turbulent flux models based on the gradient-transport approximation fail to reproduce the turbulent transport of the mass, momentum, and heat in the stellar interiors. This suggests that the buoyancy-dominated flows of core-collapse convection are characterised by large-scale plumes. The entrainment between the rising and sinking plumes strongly affects the transport properties of neutrino-driven turbulence. In particular, the turbulent dissipation, which is balanced by the buoyancy drive, is dominantly controled by the entrainment of negatively buoyant plumes. 

 It has been theoretically and experimentally shown that the dissipation rates in jets and plumes follow the Kolmogorov equilibrium law only if the spreading rates of them do not vary with the streamwise distance.\citep{lay2018,caf2019} Adopting the Lie symmetry group theory on the statistical turbulent model equations, \citet{sun2021} recently argued the non-equilibrium turbulent dissipation in buoyant axisymmetric plumes in unstratified stagnant ambient at the infinite-Reynolds-number limit. They considered the relationship between the dissipation rates of the turbulent kinetic energy and turbulent thermal fluctuation, and the entrainment coefficients in the Kolmogorov and non-Kolmogorov regions of turbulent dissipations. They showed that the well-known link between entrainment and dissipation in the Kolmogorov region of turbulence is also established in the non-Kolmogorov region.

As mentioned above, with the aid of the response-function formalism, the non-equilibrium properties of plume motions can be expressed by the variations of the turbulent kinetic energy and its dissipation rate along the plume motions. Recent elaborated laboratory experiments of jets and negatively or positively buoyant plumes provide detailed data of spatial distributions of the turbulent energy and its dissipation rate as well as the jet and plume velocity distributions.\citep{cha2017,lai2019a,lai2019b,bor2020} Utilizing these data, we can evaluate the variation of the turbulent energy and its dissipation rate along the jet/plume motion. Such an evaluation enables us to examine how much non-equilibrium effect may present in turbulent flows associated with the jet and plume. This is one of the novel points discussed in this work.
	
Negatively or positively buoyant plumes play a dominant role in mass, momentum and heat mixing in the stellar convection zone. In the stellar convection zone case, we have to simultaneously treat the effects of numerous plumes on the turbulent fluxes, such as the turbulent mass flux, momentum flux, and heat or internal energy flux, in the mean-field equations. The non-equilibrium effect associated with these numerous plumes is investigated with the aid of numerical simulations of a simplified domain mimicking the stellar convection zone. A turbulence model incorporating the non-equilibrium effect into the expressions of the turbulence fluxes should be validated in comparison with the direct numerical simulations (DNSs) of the turbulent mass, momentum and heat or internal-energy fluxes. In order to properly capture the effects of plume motions, which are part of fluctuating motions in the framework of the mean-field modeling, we introduce some elaborated averaging procedures in this evaluation of the non-equilibrium turbulence model. This is a time--space double averaging procedure, and has been recently applied to a stellar convection zone.\citep{yok2022} This point will be also discussed in this paper.

Organisation of this paper is as follows. In section~\ref{sec:com_eqs}, we present the mean-field equations with the turbulence correlations, which should be expressed in terms of the mean-field and transport coefficients representing the statistical properties of turbulence. In section~\ref{sec:noneq_effect}, closure models for the turbulent fluxes are argued with the aid of the theoretical derivations of them, with special reference to the corrections to the conventional gradient-transport approximations. There, modifications due to the non-equilibrium effects are presented. In section~\ref{sec:experiments}, roles of the non-equilibrium effects are suggested in the context of recent plume/jet experiments. In section~\ref{sec:stellar_conv}, the role of plumes in the stellar convection is discussed in the context of the stellar convection. In addition, the interaction between the coherent and incoherent fluctuations is argued using the transport equations of each component of fluctuation energy. Concluding remarks are given in section~\ref{sec:concl}.

\section{Equations of compressible hydrodynamic turbulence \label{sec:com_eqs}}\label{sec:equations}
Material of this work is plumes in compressible convective turbulence. The turbulent fluxes, such as the turbulent mass flux, the Reynolds stress, the turbulent heat or internal-energy flux, in the mean-field equations are analysed with the aid of the two-scale direct-interaction approximation (TSDIA) formulation. In this method, the response or Green's function is used to incorporate the non-equilibrium properties of the turbulence along the mean or grid-scale (GS) flow into the turbulent transport coefficients. In applying the present formulation to the statistical analysis of the turbulent transports in the presence of numerous plumes, a time--space double averaging procedure is adopted. Since there is no general way to determine the time-domain filtering, we have to make use of the physical considerations on the plume statistics and dynamics including its lifetime, for determining the filter in the time domain. This point should be further explored beyond the present formulation described below. 

\subsection{Mean-field equations and turbulent correlations}
We adopt the conventional Reynolds decomposition procedure 
\begin{equation}
	f = F + f',\;\; F = \langle {f} \rangle
	\label{eq:reynolds_ave}
\end{equation}
with
\begin{subequations}
\begin{equation}
	f = (\rho, {\bf{u}}, p, e)
	\label{eq:inst_f_def}
\end{equation}
\begin{equation}
	F = (\langle {\rho} \rangle, {\bf{U}}, P, E),
	\label{eq:mean_f_def}
\end{equation}
\begin{equation}
	f = (\rho', {\bf{u}}', p', e'),
	\label{eq:fluct_f_def}
\end{equation}
\end{subequations}
where $\rho$ is density, ${\bf{u}}$ the velocity, $p$ the pressure, $e$ the internal energy of fluid. With the Reynolds averaging (\ref{eq:reynolds_ave}), the mean-field equations are obtained in the following form; the density equation:
\begin{equation}
    \frac{\partial \langle{\rho}\rangle}{\partial t}
    + {\boldsymbol{\nabla}} {\boldsymbol{\cdot}} \left( {
        \langle{\rho}\rangle {\bf{U}}
    } \right)
    = - {\boldsymbol{\nabla}} {\boldsymbol{\cdot}} \left\langle {
	    \rho' {\bf{u}}'
    } \right\rangle,
    \label{eq:mean_den_eq}
\end{equation}
the mean velocity equation:
\begin{eqnarray}
    \frac{\partial}{\partial t} \langle{\rho}\rangle U^\alpha
    + \frac{\partial}{\partial x^a} \langle{\rho}\rangle U^a U^\alpha
    &=& - \left( {\gamma - 1} \right) 
        \frac{\partial}{\partial x^\alpha} \langle{\rho}\rangle E
    + \frac{\partial}{\partial x^a} \mu {\cal{S}}^{a \alpha}
    \nonumber\\
    &-& \frac{\partial}{\partial x^a} \left( {
	    \langle{\rho}\rangle 
    	\left\langle {u'{}^a u'{}^\alpha} \right\rangle
	    + U^a \left\langle {\rho' u'{}^\alpha} \right\rangle
        + U^\alpha \left\langle {\rho' u'{}^a} \right\rangle
    } \right)
    + \langle{\rho}\rangle g^\alpha,
    \label{eq:mean_mom_eq}
\end{eqnarray}
and the mean internal energy equation:
\begin{eqnarray}
    \frac{\partial}{\partial t} \langle{\rho}\rangle E
    + {\boldsymbol{\nabla}} {\boldsymbol{\cdot}} \left( {
	    \langle{\rho}\rangle {\bf{U}} E
    } \right)
    &=& {\boldsymbol{\nabla}} {\boldsymbol{\cdot}} \left( {
	    \frac{\kappa}{C_V} {\boldsymbol{\nabla}} E
    } \right)
    - {\boldsymbol{\nabla}} {\boldsymbol{\cdot}} \left( {
	    \langle{\rho}\rangle \left\langle {e' {\bf{u}}'} \right\rangle
	    + E \left\langle {\rho' {\bf{u}}'} \right\rangle
	    + {\bf{U}} \left\langle {\rho' e'} \right\rangle
    } \right)\nonumber\\
    && \hspace{-0pt} - \left( {\gamma - 1} \right) 
    \left( {
	    \langle{\rho}\rangle E {\boldsymbol{\nabla}} {\boldsymbol{\cdot}} {\bf{U}}
	    + \langle{\rho}\rangle \left\langle {
		    e' {\boldsymbol{\nabla}} {\boldsymbol{\cdot}} {\bf{u}}'
		} \right\rangle
	    + E \left\langle {\rho' {\boldsymbol{\nabla}} {\boldsymbol{\cdot}} {\bf{u}}'} \right\rangle
	} \right).
	\label{eq:mean_int_en_eq}
\end{eqnarray}
In (\ref{eq:mean_mom_eq}), ${\cal{S}}^{\alpha\beta}$ is the deviatric part of the mean velocity strain defined by
\begin{equation}
    {\cal{S}}^{\alpha\beta}
    = \frac{\partial U^\beta}{\partial x^\alpha}
    + \frac{\partial U^\alpha}{\partial x^\beta}
    - \frac{2}{3} {\boldsymbol{\nabla}} {\boldsymbol{\cdot}} {\bf{U}} \delta^{\alpha\beta},
    \label{eq:mean_vel_strain}
\end{equation}
and ${\bf{g}} = \{ {g^\alpha} \}$ is the gravitational acceleration.

In the mean-field equations (\ref{eq:mean_den_eq})-(\ref{eq:mean_int_en_eq}), we have several second-order turbulence correlations. They are the turbulent mass flux: $\langle {\rho' {\textbf{u}}'} \rangle$, the Reynolds stress: $\langle {{\textbf{u}}'{\textbf{u}}'} \rangle$, the turbulent internal-energy flux: $\langle {e' {\textbf{u}}'} \rangle$, the internal-energy dilatation correlation: $\langle {e' {\boldsymbol{\nabla}} {\boldsymbol{\cdot}} {\textbf{u}}'} \rangle$, the density dilatation correlation: $\langle {\rho' {\boldsymbol{\nabla}} {\boldsymbol{\cdot}} {\textbf{u}}'} \rangle$, and the density--internal-energy correlation: $\langle {\rho' e'} \rangle$. In the conventional turbulence modeling approach, the expressions of these turbulent correlations are modelled in a heuristic manner. In marked contrast, in a self-consistent turbulence modeling approach based on an analytical theory, the expressions of these correlations are obtained from a closure theory of inhomogeneous turbulence. In the present work, we use the analytical results from a multiple-scale renormalisation perturbation expansion method; two-scale direct-interaction approximation (TSDIA) analysis of the fluctuation equations.\citep{yos1984,yos1998,yok2020} Analysis of incompressible and strongly compressible turbulence with magnetic field using the TSDIA formalism can be seen, for example, in \citet{yok2013,yok2018a,yok2018b}.

\subsection{Simplest modeling turbulent fluxes}
In the turbulence modeling approaches, in order to reduce the huge burden of treating complexity of turbulent flows, it is often required to model the turbulent fluxes in a simplest possible manner. A heuristic approach of modeling is to adopt the gradient diffusion approximations for the turbulent fluxes appearing in (\ref{eq:mean_den_eq})-(\ref{eq:mean_int_en_eq}). There, the turbulent momentum, mass and energy fluxes are assumed to be written in the form:
\begin{equation}
    \langle {u'{}^\alpha u'{}^\beta} \rangle_{\textrm{D}}
    = - \nu_{\textrm{T}} {\cal{S}}^{\alpha\beta},
    \label{eq:conv_uu_model}
\end{equation}
\begin{equation}
    \langle {\rho' {\textbf{u}}'} \rangle
    = - \frac{\nu_{\textrm{T}}}{\sigma_\rho} {\boldsymbol{\nabla}} \langle{\rho}\rangle,
    \label{eq:conv_rho_u_model}
\end{equation}
\begin{equation}
    \langle {e' {\textbf{u}}'} \rangle
    = - \frac{\nu_{\textrm{T}}}{\sigma_e} {\boldsymbol{\nabla}} E,
    \label{eq:conv_qu_model}
\end{equation}
where the suffix $\textrm{D}$ denotes the deviatoric or traceless part of a tensor as ${\cal{A}}^{\alpha\beta}_{\textrm{D}} = {\cal{A}}^{\alpha\beta} - (1/3) {\cal{A}}^{aa} \delta^{\alpha\beta}$. Here, the transport coefficient $\nu_{\textrm{T}}$ is the turbulent or eddy viscosity, and $\sigma_\rho$ and $\sigma_e$ are the turbulent Prandtl numbers for the density and internal energy, respectively. The turbulent transport coefficients represent the statistical properties of turbulence. For example, the turbulent viscosity is expressed in terms of the intensity of turbulence or turbulent energy (per mass) $K (= \langle {{\textbf{u}}'{}^2} \rangle/2)$.

In the mixing-length theory (MLT) model, the turbulent viscosity is written as
\begin{equation}
    \nu_{\textrm{T}} 
    = v \ell_{\textrm{C}}
    \label{eq:nuT_MLT}
\end{equation}
apart from the numerical factor, where $v$ is the characteristic intensity of turbulent velocity and $\ell_{\textrm{C}}$ is the characteristic length in the energy containing eddies. In practical applications of the model to the geophysical and astrophysical turbulent flows, depending on the problem we consider, the pressure or density scale height is often adopted as the mixing length. In terms of the turbulent energy $K$, equation~(\ref{eq:nuT_MLT}) is expressed as
\begin{equation}
    \nu_{\textrm{T}}
    = K^{1/2} \ell_{\textrm{C}}
    \label{eq:MLT_kol_pra}
\end{equation}
(the Kolmogorov--Prandtl expression).

This model expression for the turbulent viscosity $\nu_{\textrm{T}}$ can be obtained from consideration of the equilibrium energy spectrum of turbulence as follows. The Kolmogorov scaling of turbulence is based on the equilibrium between the production rate $P_K$ and the dissipation rate $\varepsilon_K$ of the turbulent energy as
\begin{equation}
    P_K \simeq \varepsilon_K.
    \label{eq:local_equil}
\end{equation}
Hereafter, the dissipation rate of turbulent energy, $\varepsilon_K$, is simply denoted as $\varepsilon$. Assuming the equilibrium scaling of Kolmogorov:
\begin{equation}
    E(k) = Ko\ \varepsilon^{2/3} k^{-5/3}
    \label{eq:kol_scaling}
\end{equation}
[$Ko\ (\simeq 1.5)$: Kolmogorov constant], the turbulent energy is expressed as
\begin{equation}
    K = \int dk E(k)
    = \int_{k=k_{\textrm{C}}}^{\infty} dk 
        Ko\ \varepsilon^{2/3} k^{-3/5} 
    = C_{K0} \varepsilon^{2/3} \ell_{\textrm{C}}^{2/3},
    \label{eq:K_kolmogorov}
\end{equation}
where $\ell_{\textrm{C}} (= 2\pi/k_{\textrm{C}})$ is the largest eddy size. Here, we simply related it to the infrared cut-off wavenumber $k_{\textrm{C}}$, and $C_{K0}$ is the model constant linked to the Kolmogorov constant $Ko$. With (\ref{eq:MLT_kol_pra}) the turbulent or eddy viscosity $\nu_{\textrm{T}}$ in (\ref{eq:conv_uu_model})-(\ref{eq:conv_qu_model}) is evaluated as
\begin{equation}
    \nu_{\textrm{T}}
    = K^{1/2} \ell_{\textrm{C}}
    = \varepsilon^{1/3} \ell_{\textrm{C}}^{4/3}
    = K^2 / \varepsilon
    \label{eq:nuT_K_ell_eps}
\end{equation}
apart from the numerical factors. The final equality is the well-known eddy-viscosity expression in the standard $K-\varepsilon$ model. With this turbulent viscosity expression, equation~(\ref{eq:conv_uu_model}) constitutes the eddy-viscosity representation of the Reynolds stress or turbulent momentum transport.

\section{Non-equilibrium effect \label{sec:noneq_effect}}
\subsection{Non-equilibrium effect on eddy viscosity \label{sec:noneq_eddy_visc}}
\begin{figure}[h]
\centering
	\includegraphics[scale=0.7]{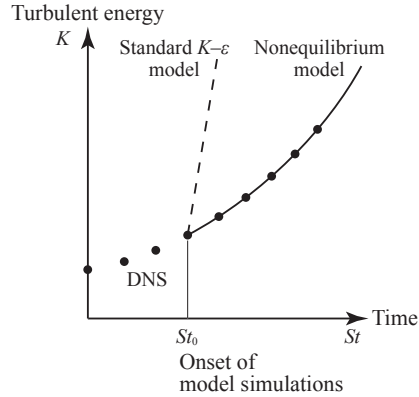}
    \caption{Schematically depicted evolution of the turbulent energy in the simulations of homogeneous shear flow. $\bullet$: DNS, $-\ -$: standard $K-\varepsilon$ model simulation, $\rule{5.ex}{0.2ex}$ : non-equilibrium model simulation. Onset of the model simulations is set at a scaled time $St_0$ using the DNS data ($S$: shear rate).}
    \label{fig:homo_shear_sims}
\end{figure}

In the standard $K-\varepsilon$ model, the turbulent viscosity is expressed in terms of the turbulent energy $K$ and its dissipation rate $\varepsilon$ as
\begin{equation}
    \nu_{\textrm{T}}
    = C_\nu \frac{K^2}{\varepsilon},
    \label{eq:nuT_K_eps}
\end{equation}
where $C_\nu$ is model constant usually taken as $C_\nu = 0.09$. However, it has been long known that the standard $K-\varepsilon$ model with the turbulent viscosity (\ref{eq:nuT_K_eps}) applied to practical flows has some obvious drawbacks. The representative situations where the standard $K-\varepsilon$ model fails include the cross-flow configuration with the circumferential or rotating velocity component as well as the axial one, impinging flow configuration. Another example is the overestimate of the turbulent energy $K$ and its dissipation rate $\varepsilon$ in the homogeneous shear turbulence. The results of the standard $K-\varepsilon$ show a tremendous overestimate of the turbulent energy $K$ as compared with the counterpart in the direct numerical simulations (DNSs) as schematically depicted in Figure~\ref{fig:homo_shear_sims}. Even in one of the simplest flow geometries such as the homogeneous shear, a simple turbulent viscosity model fundamentally fails to reproduce turbulent flow characteristics.

In order to rectify such drawbacks of the standard $K-\varepsilon$ model, several improvements of the model have been proposed. A possible improvement of the model is the implementation of non-equilibrium effect on the basis of a statistical theory of inhomogeneous turbulence. Non-equilibrium effect on turbulent viscosity has been investigated with the two-scale direct-interaction approximation (TSDIA).\citep{yos1984,oka1994} In the framework of the TSDIA, the formal solution of the lowest-order turbulent velocity arising from the large-scale inhomogeneities in space and time is given in terms of the homogeneous isotropic basic field ${\bf{u}}_0'({\bf{k}};\tau_1)$ and the response function ${\mbox{\boldmath${G}$}}'({\bf{k}};\tau,\tau_1)$ as
\begin{eqnarray}
    u_1'{}^\alpha({\bf{k}};\tau)
    &=& \frac{\partial U^b}{\partial X^a}
  \int_{-\infty}^\tau\!\!\! d\tau_1
    G'{}^{\alpha b}({\bf{k}};\tau,\tau_1)
    u_0'{}^a({\bf{k}};\tau_1)
\nonumber\\
&& \hspace{-30pt}- \int_{-\infty}^\tau\!\!\! d\tau_1 
    G'{}^{\alpha a}({\bf{k}};\tau,\tau_1)
    \frac{Du_0'{}^a({\bf{k}};\tau_1)}{DT}
\nonumber\\
&&\hspace{-30pt}+ 2 M^{dab}({\bf{k}}) \iint \delta({\bf{k}}- {\bf{p}}-{\bf{q}}) d{\bf{p}} d{\bf{q}}
  \int_{-\infty}^\tau\!\!\! d\tau_1 
    G'{}^{\alpha d}({\bf{k}};\tau,\tau_1)
        \frac{q^b}{q^2} u_0'{}^a({\bf{p}};\tau_1)
    \frac{\partial u'{}_0^c({\bf{q}};\tau_1)}{\partial X^c}
\nonumber\\
&&\hspace{-30pt} - M^{abcd}({\bf{k}}) \iint \delta({\bf{k}}- {\bf{p}}-{\bf{q}}) d{\bf{p}} d{\bf{q}}
  \int_{-\infty}^\tau\!\!\! d\tau_1 
    G'{}^{\alpha d}({\bf{k}};\tau,\tau_1)
    \frac{\partial}{\partial X^c}
    (u_0'{}^a({\bf{p}};\tau_1) 
      u_0'{}^b({\bf{q}};\tau_1)).
    \label{eq:u1_exp}
\end{eqnarray}
Here, the second term on the right-hand side related to the Lagrange derivative represents the non-equilibrium properties of turbulent field along the flow. The turbulent kinetic energy $K$ with the non-equilibrium effect is written as
\begin{eqnarray}
    K
    &=& \langle {{\bf{u}}'{}^2} \rangle/2
    = \frac{1}{2} \int d{\bf{k}}\ 
          \langle {u'{}^\ell({\bf{k}};\tau) u'{}^\ell({\bf{k}}';\tau)} \rangle / \delta({\bf{k}} + {\bf{k}}')
    \nonumber\\
    &=& (\langle {u_0'{}^\ell u_0'{}^\ell} \rangle
    + \langle {u_0'{}^\ell u_1'{}^\ell} \rangle 
    + \langle {u_1'{}^\ell u_0'{}^\ell} \rangle 
    + \cdots)/2
    \nonumber\\
    &=& C_{K1} \varepsilon^{2/3} \ell_{\textrm{C}}^{2/3}
    - C_{K2} \varepsilon^{-3/2} \ell_{\textrm{C}}^{4/3} 
        \frac{D\varepsilon}{Dt}
    - C_{K3} \varepsilon^{1/3} \ell_{\textrm{C}}^{1/3}
        \frac{D\ell_{\textrm{C}}}{Dt},
    \label{eq:K_exp_eps_ell}
\end{eqnarray}
where $\ell_{\textrm{C}}$ is the length scale of the largest eddy and $D/Dt (= \partial/\partial t + {\textbf{U}} {\boldsymbol{\cdot}} {\boldsymbol{\nabla}})$ is the Lagrange or material derivative along the mean-flow velocity. Here, $C_{Kn} (n=1-3)$ are the model constants whose magnitude are related to those of the energy spectrum and time scale of turbulence [$\sigma_0$ in (\ref{eq:sigma_scaling}) and $\omega_0$ in (\ref{eq:omega_scaling})]. A brief derivation of (\ref{eq:K_exp_eps_ell}) is given in appendix~\ref{sec:append_A}.

	We solve (\ref{eq:K_exp_eps_ell}) with respect to $\ell_{\textrm{C}}$ by an iteration method with the first term of (\ref{eq:K_exp_eps_ell}) as the starting term:
\begin{equation}
    K = C_{K1} \varepsilon^{2/3} \ell_{\textrm{C}}^{2/3}.
    \label{eq:K0_exp_eps_ell}
\end{equation}
Then we have $\ell_{\textrm{C}}$ as
\begin{equation}
    \ell_{\textrm{C}} = C_{\ell 1} K^{3/2} \varepsilon^{-1},
    \label{eq:ell_exp_K_eps}
\end{equation}
where $C_{\ell 1}$ is the model constant. With (\ref{eq:MLT_kol_pra}), this lowest-order expression of the length scale corresponds to the simplest eddy-viscosity model (\ref{eq:nuT_K_eps}).

Substituting (\ref{eq:nuT_K_eps}) into (\ref{eq:K_exp_eps_ell}), we have the first iteration result as
\begin{equation}
    \ell_{\textrm{C}}
    = C_{\ell 1} K^{3/2} \varepsilon^{-1}
    + C_{\ell 2} K^{3/2} \varepsilon^{-2} \frac{DK}{Dt}
    - C_{\ell 3} K^{5/2} \varepsilon^{-3} \frac{D\varepsilon}{Dt}
    \label{eq:ell_exp_K_eps_NE}
\end{equation}
($C_{\ell 2}$ and $C_{\ell 3}$ are model constants). The second and third terms of (\ref{eq:ell_exp_K_eps_NE}) represent the non-equilibrium effect. Equation~(\ref{eq:ell_exp_K_eps_NE}) shows that the length scale of turbulence, as well as the time scale and energy of turbulence, is corrected by the $D/Dt$-related non-equilibrium effect terms. Equation~(\ref{eq:ell_exp_K_eps_NE}) can be approximated as
\begin{equation}
    \ell_{\textrm{C}}
    = \ell_{\textrm{E}} \left( {
        1 - C'_{\textrm{N}} \frac{1}{K} 
        \frac{D}{Dt} \frac{K^2}{\varepsilon}
    } \right),
    \label{eq:ell_C_non_equil}
\end{equation}
where $\ell_{\textrm{E}}$ is the equilibrium characteristic length defined by
\begin{equation}
    \ell_{\textrm{E}} = K^{3/2}/\varepsilon,
    \label{eq:ellE_exp_K_eps}
\end{equation}
and $C'_{\textrm{N}}$ is the non-equilibrium-effect-related coefficient. Depending on the streamwise development of the turbulent energy $K$ and its dissipation rate $\varepsilon$, the length scale of turbulence changes as compared to the equilibrium cases. At the same time, the time scale of the non-equilibrium turbulence may be expressed as
\begin{equation}
    \tau_{\textrm{NS}}
    = \frac{\ell_{\textrm{NS}}}{K^{1/2}}
    \simeq C_{\textrm{S}}^{-3/2} \frac{K}{\varepsilon} 
        \left( {
        1 - C'_{\textrm{N}} \frac{1}{K} 
        \frac{D}{Dt} \frac{K^2}{\varepsilon}
        } \right).
    \label{eq:tauNS_nonequiv}
\end{equation}
This indicates that the effective time scale may change depending on the $D/Dt$ effect.

The non-equilibrium turbulent viscosity $\nu_{\textrm{T}}$ can be expressed as
\begin{equation}
    \nu_{\textrm{T}}
    = C_\nu \frac{K^2}{\varepsilon}
    - C_{\nu K} \frac{K^2}{\varepsilon^2} \frac{DK}{Dt}
    + C_{\nu \varepsilon} 
        \frac{K^3}{\varepsilon^3} \frac{D\varepsilon}{Dt},
    \label{eq:nuT_exp_K_eps_nonequil}
\end{equation}
where $C_\nu$, $C_{\nu K}$, and $C_{\nu\varepsilon}$ are the model constants, whose values can be evaluated from theoretical analyses. Equation~(\ref{eq:nuT_exp_K_eps_nonequil}) is rewritten as
\begin{equation}
    \nu_{\textrm{T}}
    = C_\nu \frac{K^2}{\varepsilon} \left[ {
        1 - \left( {
            C_{{\textrm{N}} K} \frac{1}{\varepsilon} \frac{DK}{Dt}
            - C_{{\textrm{N}} \varepsilon} 
                \frac{K}{\varepsilon^2} \frac{D\varepsilon}{Dt}  
            } \right)
    } \right],
    \label{eq:nuT_exp_DK_Deps}
\end{equation}
or approximately,
\begin{equation}
    \nu_{\textrm{T}}
    = C_\nu \frac{K^2}{\varepsilon} \left( {
        1 - C_{\textrm{N}} 
        \frac{1}{K} \frac{D}{Dt}\frac{K^2}{\varepsilon}
    } \right)
    \label{eq:nuT_exp_nuN}
\end{equation}
($C_{{\textrm{N}}K}$, $C_{{\textrm{N}}\varepsilon}$, and $C_{\textrm{N}}$: model constants). Here we should note that this non-equilibrium effect does not necessarily require the compressibility. This effect shows up even in the incompressible cases.

In practical situations in simulations of turbulent flows, a large positive $D/Dt$ term may occur locally, in such a case, (\ref{eq:nuT_exp_nuN}) may lead to a physically undesirable negative turbulent viscosity. In order to alleviate such an unphysical model result to enhance the applicability of the model expression, we use a simple Pad\'{e} approximation to (\ref{eq:nuT_exp_nuN}) and have
\begin{equation}
    \nu_{\textrm{T}}
    = \left\{ {
    \everymath{\displaystyle}
    \begin{array}{lll}
    \nu_{\textrm{TE}} \left( {
      1 + C_{\textrm{N}} 
            \frac{1}{K} \frac{D}{Dt}\frac{K^2}{\varepsilon}
    } \right)^{-1}
    &\hspace{10pt} \mbox{for} \hspace{10pt}
    &\frac{D}{Dt} \frac{K^2}{\varepsilon} >0,
    \\
    \rule{0.ex}{6.ex}
    \nu_{\textrm{TE}} \left( {
        1 - C_{\textrm{N}} 
            \frac{1}{K} \frac{D}{Dt}\frac{K^2}{\varepsilon}
    } \right)
    &\hspace{10pt} \mbox{for} \hspace{10pt}
    &\frac{D}{Dt} \frac{K^2}{\varepsilon} <0,
    \end{array}
    } \right.
    \label{eq:nuTN_both_cases}
\end{equation}
where the equilibrium eddy viscosity $\nu_{\textrm{TE}}$ is defined in the same form as (\ref{eq:nuT_K_eps}) by
\begin{equation}
    \nu_{\textrm{TE}}
    = C_\nu \frac{K^2}{\varepsilon}.
    \label{eq:nuTE_K0_eps0}
\end{equation}
Equation~(\ref{eq:nuTN_both_cases}) indicates that, depending on the sign of the Lagrangian or material derivative $(D/Dt) (K^2/\varepsilon)$, the turbulent viscosity is effectively decreased or increased by the non-equilibrium effect. In the case of $(D/Dt) (K^2/\varepsilon) >0$, $\nu_{\textrm{T}}$ is decreased, while in the case of $(D/Dt) (K^2/\varepsilon) <0$, $\nu_{\textrm{T}}$ is increased. This implies that in this case the non-equilibrium effect works in the direction to stabilise the deviation from the equilibrium state (return to equilibrium).

The non-equilibrium model expression for the eddy viscosity (\ref{eq:nuTN_both_cases}) has been applied to the homogeneous-shear turbulence, where the standard $K-\varepsilon$ model with the conventional eddy viscosity (\ref{eq:nuTE_K0_eps0}) leads to overestimates of the turbulent energy $K$ and its dissipation rate $\varepsilon$ as compared to the DNS results (Figure~\ref{fig:homo_shear_sims}). It was reported that the non-equilibrium model (\ref{eq:nuTN_both_cases}) with $C_{\textrm{N}} = 0.8$ leads to the results consistent with the DNS ones.\citep{yos1993} In this sense, the validity of the non-equilibrium effects in turbulent mixing has been already confirmed in a simple geometry flow.

\section{Non-equilibrium effects in jet and plume experiments \label{sec:experiments}}
As was referred to in section~\ref{sec:intro}, in the buoyancy-dominated turbulent flows, the turbulent convection is characterised by large-scale plumes. As this consequence, the turbulent fluxes are deviated from the usual gradient-transport approximation. It has been recognised that entrainment dominates the evolution of buoyant plumes.\citep{spr1997} As was shown in a previous study of supernova convection \citep{mur2011}, global turbulence models introducing the entrainment parameters provide the best results for describing the evolution of buoyant plumes. However, such models highly depend on the geometrical configurations of the simulation, and are difficult to apply to a general 3D situation. In this sense, it is preferable to develop a local model without resorting to global parameters, provided that the local model works well enough for reproducing the basic behaviour of the convection turbulence with plumes. In order to validate the model, here we consider turbulent flows with much simpler geometrical configurations, jets and plumes, utilising recent experimental results.

\begin{figure}[h]
\centering
	\includegraphics[scale=0.7]{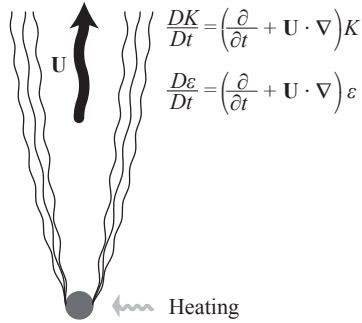}
    \caption{Non-equilibrium effect associated with a plume motion. A plume is thermobuoyantly formed above a heat source. The non-equilibrium effect is taken into account through the Lagrangian or advective derivatives of the turbulent energy $K$ and its dissipation rate $\varepsilon$, $DK/Dt = (\partial/\partial t + {\bf{U}} \cdot \nabla) K$ and $D\varepsilon/Dt = (\partial/\partial t + {\bf{U}} \cdot \nabla) \varepsilon$, where ${\bf{U}}$ is the plume velocity.}
    \label{fig:nonequil_plume}
\end{figure}

The non-equilibrium effect may help to develop such a local model. In the presence of a plume, we can expect that the turbulence properties vary along the plume-flow direction. The non-equilibrium effect associated with the plume motion can be incorporated through the variation of the turbulent energy $K$ and its dissipation rate $\varepsilon$ along the plume motion in terms of the Lagrangian or advective derivatives, $DK/Dt = [(\partial/ \partial t) + {\bf{U}} \cdot \nabla] K$ and $D\varepsilon/Dt = [(\partial/ \partial t) + {\bf{U}} \cdot \nabla] \varepsilon$ (Figure~\ref{fig:nonequil_plume}). The effect of non-equilibrium turbulence can be implemented into the model of turbulent viscosity as (\ref{eq:nuTN_both_cases}) with the equilibrium turbulent viscosity given by (\ref{eq:nuTE_K0_eps0}).

Recently, a series of detailed experimental studies of the budgets of turbulent energy, Reynolds stresses, and dissipation rate have been performed in single- and multi-phase jets/plumes, including a turbulent round jet discharged into an initially stationary ambient \citep{lai2019a}, a buoyant multi-phase bubble plume \citep{lai2019b}, a negatively buoyant multi-phase particle plume \citep{bor2020}, and a variable-density round jet with co-flow.\citep{cha2017}

\subsection{Round jets\label{sec:round_jets}} 

Using a set of stereoscopic particle image velocimetry (SPIV) measurement of a turbulent round water jet, budget of the turbulent energy equation was investigated in \citet{lai2019a}. In this experiment, it was reported that the magnitudes of the normal Reynolds stresses $R^{ii}$ decrease by 30 $\%$ over $27 \le x/D \le 37$ in the downstream direction ($D$: exit diameter of the jet nozzle). In particular, the streamwise variation of the axial velocity fluctuation $\langle {(u'{}^x)^2} \rangle / \sqrt{\varepsilon}$ at the jet centreline shows a decreasing tendency at the major region of the axial locations $31 \le x/D \le 37$ (Table~\ref{tab:tab1}). Although in \citet{lai2019a} this experimental results are argued in the context of the approximate achievement of self-preserving state in the jet \citep{ten1972}, this decrease tendency may be interpreted as a deviation from the self-preserving state due to the non-equilibrium effect. Since the magnitude of axial velocity fluctuation $\sqrt{\langle {(u'{}^x)^2} \rangle}$ is similar to and just slightly larger than the horizontal counterparts of the velocity fluctuations, $\sqrt{\langle {(u'{}^y)^2} \rangle}$ and $\sqrt{\langle {(u'{}^z)^2} \rangle}$, and since the energy dissipation rate $\varepsilon$ is well represented by that of the isotropic turbulence, $\varepsilon_{\textrm{iso}}$, the evolution of squared axial velocity fluctuation, $\langle {(u'{}^x)^2} \rangle / \sqrt{\varepsilon_{\textrm{iso}}}$, provides a reasonable surrogate for the basic behaviour of $K^2 / \varepsilon$:
\begin{equation}
    \frac{K^2}{\varepsilon}
    \propto \left[ {
    \frac{\langle {({u'{}^x})^2} \rangle}{\sqrt{\varepsilon_{\textrm{iso}}}}
    } \right]^2.
    \label{eq:sqr_vel_fluct}
\end{equation}
\begin{table}[h] 
\caption{
Streamwise variation of normalised turbulent intensities, turbulent energy, and turbulent axial velocity fluctuation normalised by the square root of the dissipation rate at the jet centreline.\label{tab:tab1} $x/D$: jet streamwise length, $D$: jet nozzle exit diameter ($D=11\ {\textrm{mm}}$), $U_{\textrm{c}}$: centreline jet velocity ($U_{\textrm{c}} = 50\ {\textrm{mm}}\ {\textrm{s}}^{-1}$). Data are taken from figures 7 and 8 of \citet{lai2019a}.
}
\begin{center}
\begin{tabular}{cccccccc}
{$x/D$}	& $\sqrt{\langle {(u'{}^x)^2} \rangle} / U_{\textrm{c}}$	& $\sqrt{\langle {(u'{}^y)^2} \rangle} / U_{\textrm{c}}$    & $\sqrt{\langle {(u'{}^z)^2} \rangle} / U_{\textrm{c}}$    & $K / U_{\textrm{c}}^2$   &$\langle {(u'{}^x)^2} \rangle / \sqrt{\varepsilon_{\textrm{iso}}}$  &\\
31	& 0.26	& 0.20  & 0.19 & 0.1437  & $3.25\ \times 10^{-2}$\\
37    & 0.26 & 0.20  & 0.19 & 0.1437 & $2.65\ \times 10^{-2}$\\
    &   &	&   &   &(-20\%)\\
\end{tabular}
\end{center}
\end{table}

	The decrease of $\langle {(u'{}^x)^2} \rangle / \sqrt{\varepsilon_{\textrm{iso}}}$ in the downstream direction indicates that the sign of the non-equilibrium effect term is negative as
\begin{equation}
    \frac{D}{Dt} \frac{K^2}{\varepsilon} <0\;\;\;
    (\mbox{in turbulent round jets}).
    \label{eq:noneq_fctr_rnd_jet}
\end{equation}
It follows from (\ref{eq:nuTN_both_cases}) that the turbulent viscosity is enhanced by the non-equilibrium effect as
\begin{equation}
    \nu_{\textrm{NE}} 
    = \nu_{\textrm{E}} \left( {
        1 
        - C_{\textrm{N}} \frac{1}{K}
            \frac{D}{Dt} \frac{K^2}{\varepsilon}
    } \right)
    > \nu_{\textrm{E}}.
    \label{eq:nu_NE_round_jet}
\end{equation}.

In order to evaluate the non-equilibrium effect, we introduce the non-equilibrium correction factor $\Lambda$ by
\begin{equation}
    \Lambda 
    = C_{\textrm{N}} \frac{1}{K}
        \frac{D}{Dt} \frac{K^2}{\varepsilon},
    \label{eq:Lambda_def}
\end{equation}
where $C_{\textrm{N}}$ is the model constant estimated as the order of unity ($C_{\textrm{N}} = 0.8$) through applications to the homogeneous-shear flow turbulence.\citep{yos1993} This factor is approximated in terms of the Lagrangian derivative as
\begin{equation}
    \Lambda 
    \simeq C_{\textrm{N}} \frac{1}{K} 
        ({\textbf{U}} {\boldsymbol{\cdot}} {\boldsymbol{\nabla}}) \frac{K^2}{\varepsilon}.
     \label{eq:Lambda_approx}
\end{equation}
Using data of \citet{lai2019a}, $\Lambda$ is estimated as
\begin{eqnarray}
	\Lambda
	\simeq C_{\textrm{N}} 
		\left( {
			\frac{U_{\textrm{c}}}{\sqrt{\langle {(u'{}^x)^2} \rangle}}
		} \right)^2
		\frac{1}{U_{\textrm{c}}}
		\frac{\Delta\{[\langle {(u'{}^x)^2} \rangle 
				/ \sqrt{\langle {\varepsilon_{\textrm{iso}}} \rangle}]^2 \}}
			{D \Delta(x/D)}
	\simeq 1.7
	\label{eq:Lambda_round_jets}
\end{eqnarray}
with $C_{\textrm{N}} = 1$ (Table~\ref{tab:tab1}). Here, $\Delta \{ [\langle {(u^x)^2} \rangle / \sqrt{\langle {\varepsilon_{\textrm{iso}}} \rangle}]^2 \}$ in the nominator and $\Delta(x/D)$ in the denominator are the increments of $[\langle {(u^x)^2} \rangle / \sqrt{\langle {\varepsilon_{\textrm{iso}}} \rangle}]^2$ and $x/D$ along the jet streamwise direction.

In \citet{lai2019a}, the budgets of the turbulent-energy and dissipation-rate equations were also investigated. Using the balance in the budgets, the coefficients of each term of the $K-\varepsilon$ model were examined. They suggested the befitting directions for the change of model coefficients. The eddy viscosity constant $C_\nu$ has to be increased from the standard value of $C_\nu = 0.09$ to $ 0.09-0.27$, whereas a reduction of $C_\nu$ to $0.07$ is suggested from the matching for the jet spreading rate. This implies that it is inadequate to use a constant model coefficient $C_\nu$ for the turbulent viscosity. Enhancement of eddy viscosity due to the non-equilibrium effect is basically in a correct direction to the evaluation of $C_\nu$, and the estimated value of $\Lambda$ (\ref{eq:Lambda_round_jets}) is consistent with the suggested change of $C_\nu$. The degree of enhancement/suppression should be determined by the non-equilibrium property of the turbulence, which depends on the spatial location.

\subsection{Buoyant bubble plumes
\label{sec:buoy_bubble_plumes}}

In \citet{lai2019b}, the turbulent energy budgets in bubble plume were experimentally investigated. There, the turbulent energy evolution was measured both in the adjustment phase of the plume dependent on the source conditions (source size and geometry) and in the asymptotic phase independent of the source condition.

\begin{table}[h]
\caption{
Evolutions of turbulent energy and its dissipation rate along the plume flow.\label{tab:tab2} Case A is the asymptotic phase and and Case B is the adjustment phase. $z/D$: vertical streamwise height, $D$: dynamic length scale ($D=6.8\ {\textrm{cm}}$ for Case A and $D=20.4\ {\textrm{cm}}$ for Case B), $W_{\textrm{c}}$: plume centreline velocity, and $b_g$: Gaussian plume radius. Data are taken from figures 20 and 25 of \citet{lai2019b}.
}
\begin{center}
Case A (asymptotic phase)\\
\begin{tabular}{cccccccc}
{$z/D$}	& {$W_{\textrm{c}}$ ($\textrm{
cm}\ {\textrm{s}}^{-1}$)}	& {$b_g$}    & $K/W_{\textrm{c}}^2$    &$(\varepsilon/W_{\textrm{c}}^3) b_g \times 10^3$   &$K$ (${\textrm{cm}}^2\ {\textrm{s}}^{-2}$)  &$\varepsilon$ (${\textrm{cm}}^2\ {\textrm{s}}^{-3}$) &$K^2/\varepsilon$ (${\textrm{cm}}^2\ {\textrm{s}^{-1}}$)\\
6.6	& 20.64	& 5.46  & 0.135 & 0.04  & 57.51 & 64.42 & 51.34\\
11.0    & 17.90 & 8.00  & 0.165 & 0.086 & 52.87& 61.65 & 45.34\\
    &   &   &   &   &(-8.1\%)   &(-4.3\%)   &(-11.7\%)\\
\end{tabular}
\vspace{10pt}

Case B (adjustment phase)\\
\begin{tabular}{cccccccc}
{$z/D$}	& {$W_{\textrm{c}}$ ($\textrm{
cm}\ {\textrm{s}}^{-1}$)}	& {$b_g$}    & $K/W_{\textrm{c}}^2$    &$(\varepsilon/W_{\textrm{c}}^3) b_g \times 10^3$   &$K$ (${\textrm{cm}}^2\ {\textrm{s}}^{-2}$)  &$\varepsilon$ (${\textrm{cm}}^2\ {\textrm{s}}^{-3}$) &$K^2/\varepsilon$ (${\textrm{cm}}^2\ {\textrm{s}}^{-1}$)\\
2.19    & 28.50 & 5.27  & 0.15  & 0.035     & 121.84    & 153.74    & 96.56\\
3.66    & 25.14 & 8.53  & 0.225 & 0.0896    & 142.20    & 165.78 & 121.97\\
    &   &   &   &   & (+16.7 \%) & (+7.8\%) & (+26.3 \%)\\
\end{tabular}
\end{center}
\end{table}

These experimental results show that in the asymptotic phase (Case A) the turbulent energy $K$, its dissipation rate $\varepsilon$, and $K^2/\varepsilon$ in the plume core region ($r \le b_g$) decrease in the axial or downstream direction along the plume velocity, while in the adjustment phase (Case B) $K$, $\varepsilon$, and $K^2/\varepsilon$ increase (Table~\ref{tab:tab2}). However, the relative variation of turbulent kinetic energy at the downstream to the upstream is much higher in the adjustment phase (Case B) than in the asymptotic phase (Case A). This means that the turbulent energy inhomogeneity along the plume flow, which may lead to the non-equilibrium effect, is more prominent in the adjustment phase than in the asymptotic phase. This is natural since the non-equilibrium property is expected to be much higher in the adjustment phase than in the asymptotic phase. In the adjustment phase, the Lagrangian derivative of $K^2/\varepsilon$ is positive as
\begin{equation}
    \frac{D}{Dt} \frac{K^2}{\varepsilon} >0\;\;\;
    (\mbox{in the adjustment phase of buoyant bubble plume}).
    \label{eq:bubble_plume}
\end{equation}
In this case, from (\ref{eq:nuTN_both_cases})
\begin{equation}
    \nu_{\textrm{NE}}
    = \nu_{\textrm{E}} \left( {
        1 + \Lambda
    } \right)^{-1}
    < \nu_{\textrm{E}}\;\;\;
    (\mbox{in the adjustment phase of buoyant bubble plume}).
    \label{eq:nu_NE_in_chi}
\end{equation}
In the experiment of \citet{lai2019b}, $\Lambda$ is estimated as
\begin{eqnarray}
    \Lambda
    &\simeq& C_{\textrm{N}} \frac{1}{K} W_{\textrm{c}} \frac{\Delta(K^2/\varepsilon)}{D \Delta (z/D)}
    \simeq 0.20
    \label{eq:Lambda_buoy_plume}
\end{eqnarray}
with $C_{\textrm{N}} = 1$ (Table~\ref{tab:tab2}). Here $\Delta (K^2/\varepsilon)$ and $\Delta (z/D)$ are the increments of $K^2/\varepsilon$ and $z/D$ along the vertical streamwise height, respectively. This suggests that due to the non-equilibrium effect, the turbulent mixing in the buoyant bubble plumes is expected to be suppressed by a few 10\% as compared to the standard equilibrium turbulent transport.

\subsection{Variable density jets
\label{sec:var_den_jets}}
There are several studies on the buoyancy driven variable-density turbulence with jets and plumes.\citep{liv2008,cha2017} In recent experiments by \citet{cha2017}, variable density effects in turbulent round jets with coflow at high and low density ratios are experimentally investigated. There the axial and radial mixing mechanisms of jet are compared between the cases with low and high variable density contrast. It was confirmed that the Reynolds stress is suppressed by the turbulent mass flux in the increased variable density case. It was also reported that the radial transport of momentum and the energy cascade down to small scales are significantly suppressed in the variable density case. These suppressions take place dominantly in locations where the density fluctuation is large.

At the upstream region ($x_1/d_0 <16.3$, $x_1$: axial location, $d_0$: inner diameter), the turbulent energy, irrespective of whether averaging is mass-weighted or the Reynolds, and whether turbulent energy is normalised by excess momentum or the initial kinetic energy of the mean flow, develops with downstream distance along the centreline as
\begin{equation}
	\frac{DK}{Dt} > 0\;\;\;
    (\mbox{at upstream region of variable density jets}).
	\label{eq:var_dens_DKDT}
\end{equation}
This increase in turbulent energy along the mean flow takes place both in the small and large density variation cases. Since the evolution of dissipation rate $\varepsilon$ with the axial distance is not explicitly presented in \citet{cha2017}, that of $K^2/\varepsilon$ cannot be accurately evaluated. However, the analysis of several flows in \S~\ref{sec:neq_stell_conv} implies that the trend of $K^2/\varepsilon$ evolution along the mean flow or the sign of $K^2/\varepsilon$ can be surmised from the evolution of $K$ itself. From the non-equilibrium effect $D(K^2/\varepsilon)/Dt [\sim (K/\varepsilon) DK/Dt]$ in (\ref{eq:nuTN_both_cases}), we see that the turbulent transport is suppressed at the upstream region.

At the same time, \citet{cha2017} showed that the variable density effects lower both the peak and asymptotic values of turbulent energy. By considering the budget in the evolution equation of turbulent energy, they suggested that in the high variable density case, density fluctuation contribute to the suppression of energy transfer into small-scale fluctuations and preserving the mean-flow structures further downstream. These features can be well understood from the viewpoint of density-variance effect; the presence of density variance $K_\rho (= \langle {\rho'{}^2} \rangle)$ leads to suppression of turbulent transport as in (\ref{eq:nuTN_both_cases}). \citet{cha2017} concluded that the variable density effect must be modelled to accurately capture dynamical mixing in jets and other variable density flow phenomena. The combination of the non-equilibrium and density-variance effects proposed in the present work is a candidate for modeling the variable-density effect in these flows.

\section{Plumes in stellar convection zone\label{sec:stellar_conv}}
In the previous section, we argued the noneqiulibrium effect associated with a single plume and jet. In modeling such a situation, the velocity of the single plume/jet is treated as the local mean velocity, and the non-equilibrium effect associated with the mean velocity can be captured by the Lagrangian or advective derivative $D/Dt = \partial/\partial t + {\bf{U}} \cdot \nabla$ [${\bf{U}} (\equiv \langle {{\bf{u}}} \rangle)$: mean velocity]. However, in the practical application of the turbulence model to geo/astrophysical flow phenomena, we often encounter situations where the effect of numerous plumes and jets on the global or averaged flow dynamics have to be properly evaluated.

One example is the turbulent transport in the stellar convection zone. In some flow configurations, such as in the convective motion in a closed domain, the mean velocity formulated with a simple ensemble average or averaging in the statistically homogeneous plane is very small because of the statistical cancellations due to symmetry. Also such an averaged quantity cannot capture localized flow structures like plumes. Because of the small mean flow ($\langle {{\textbf{u}}} \rangle \equiv {\textbf{U}} \simeq 0$), the non-equilibrium effect represented by the mean flow advection, $({\textbf{U}} {\boldsymbol{\cdot}} {\boldsymbol{\nabla}}) K^2 / \varepsilon$ in $DK/Dt$, is not suitable for capturing the plume effects. With a simple ensemble or space averaging procedure, the direct impact due to the non-equilibrium effect might be negligible. However, the presence of grid-scale (GS) or large-eddy velocity component suggests that the counterpart of the non-equilibrium effect at GS may arise as a result of the alternation of the subgrid-scale (SGS) turbulent transport.

\subsection{Non-equilibrium effect in the stellar convection\label{sec:neq_stell_conv}}
\begin{figure}[h]
\centering
	\includegraphics[scale=0.9]{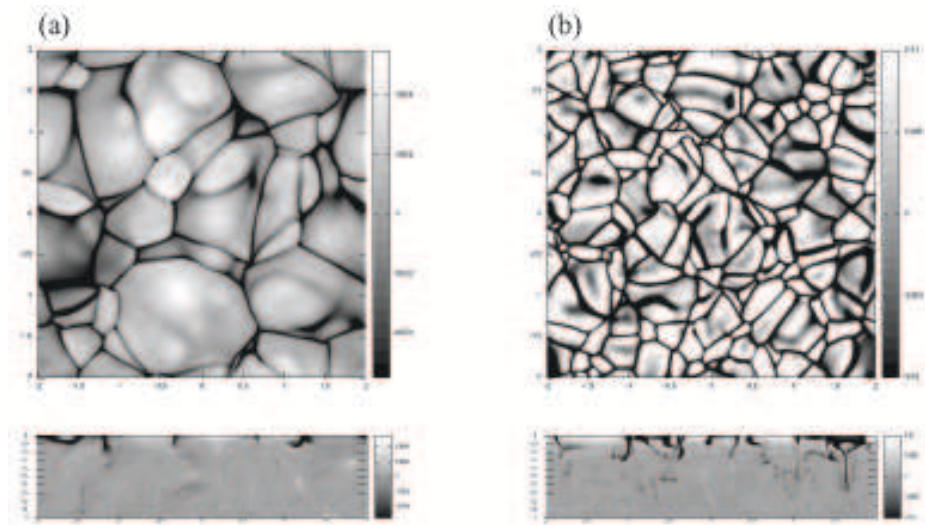}
    \caption{Entropy distributions in the direct numerical simulations (DNSs) for the locally-driven case (a) and the non-locally driven case (b). The horizontal cross-sections of the entropy fluctuation $s' (= s - \langle{s}\rangle)$ at the top surface (Top). In the non-locally- or cooling-driven case, the horizontal extension of the cell structures is much more limited than the counterpart in the locally-driven case. The vertical cross-sections of the entropy fluctuation $s'$ from the horizontal mean (Bottom). In the non-locally driven case, the low entropy down-flow or plume structures produced at the surface are prominent in the upper region.}
    \label{fig:entropy_contour_dns}
\end{figure}

Recently, in the context of the stellar convection, the non-equilibrium effects associated with the convective plume motion have been investigated in the framework of double averaging.\citep{yok2022} For the purpose of exploring the non-equilibrium effect associated with the plumes, fluid dynamics in the stellar convection zone was investigated with the aid of direct numerical simulations (DNSs). In the setup of the simulations, the diving plumes are driven by the cooling layer at the upper surface of the stellar convection zone. Because of the cooling the surface layer is convectively unstable. 
In the non-locally-driven convection case, there exist a lot of downward plumes in the shallow region of the convection zone as in the lower plot of Figure~\ref{fig:entropy_contour_dns}(b).
	
In order to capture the dynamics of plume motion, the time--space double averaging procedure was adopted there. A field quantity $f$ is divided into
\begin{equation}
    f 
	= \overbrace{
		\langle {\overline{f}} \rangle
		+ \underbrace{\tilde{f}}_
	{
	\begin{array}{c}
		\overline{f} - \langle {\overline{f}} \rangle\\
		\mbox{coherent}\\
		\mbox{fluctuation}
	\end{array}
	}
	}^{\overline{f}}
	+ \underbrace{f''}_{
	\begin{array}{c}
	f - \overline{f}\\
	\mbox{incoherent}\\
	\mbox{fluctuation}
	\end{array}}.
	\label{eq:time-space_doub_ave_smry}
\end{equation}
where $\overline{f}$ denotes the time average, $\langle {f} \rangle$ the space or ensemble average, $\tilde{f} (= \overline{f} - \langle {\overline{f}} \rangle)$ the residual of temporal average subtracted by the ensemble average part. This part may be called the dispersion or coherent fluctuation of $f$, $f'' (= f - \overline{f})$ the fluctuation around the time average $\overline{f}$. As for the time averaging, the resolved part of a quantity $f$, $\overline{f}$, is defined by the relation
\begin{equation}
	\overline{f}({\bf{x}};t)
	= \int f({\bf{x}};s) G(t-s) ds
	\label{eq:time_ave_def}
\end{equation}
with the simplest top-hat time filter function $G(t-s)$ defined by
\begin{equation}
    G(t-s) = \left\{ {
    \begin{array}{ll}
         1/T & (|t-s| \le T/2),\\
         0   & (\mbox{otherwise}).
    \end{array}
    } \right.
    \label{eq:tophat_filter}
\end{equation}
Here, $T$ is the averaging time window, which should be put in the range:
\begin{equation}
	\tau \ll T \ll \Xi,
	\label{eq:time_ave_window}
\end{equation}
where $\tau$ is the eddy turnover time of turbulence and $\Xi$ is the time scale of the mean-field evolution.

	In this formalism, the fluctuation around the space or ensemble average, $f'$, is divided into the coherent and incoherent parts as
\begin{equation}
	f' = \tilde{f} + f''.
	\label{eq:coh_incoh_fluct_def}
\end{equation}
With this double averaging procedure, the time averaged velocity $\overline{\bf{u}}$ is divided into space averaged velocity $\langle {\overline{\bf{u}}} \rangle$ and the dispersion velocity $\tilde{\bf{u}}$ as
\begin{equation}
	\overline{\bf{u}} 
	= \langle {\overline{\bf{u}}} \rangle 
	+ \tilde{\bf{u}}.
	\label{eq:coh_incoh_vel_fluct}
\end{equation}
The non-equilibrium effect along the plume motion is represented by the factor
\begin{equation}
	\widetilde{\Lambda}_{\rm{D}}
	= \left\langle {
		(\tilde{\bf{u}} \cdot \nabla) \overline{{\bf{u}}'{}^2}
	} \right\rangle.
	\label{eq:no-equib_factor}
\end{equation}
This means that if the turbulent energy is inhomogeneous along the plume motion $\tilde{\bf{u}}$, the turbulent fluxes are enhanced or suppressed depending on the sign of $\widetilde{\Lambda}_{\rm{D}}$.

\begin{figure}[h]
\centering
	\includegraphics[scale=0.7]{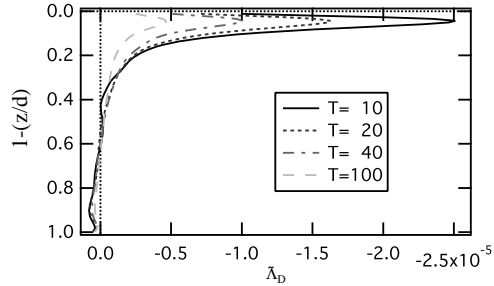}
    \caption{Spatial distributions of the non-equilibrium effect factor $\widetilde{\Lambda}_{\rm{D}} = \langle {(\widetilde{\bf{u}} \cdot \nabla) \overline{{\bf{u}}'{}^2}} \rangle$. The left axis $1- (z/d)$ is the depth from the surface ($z=d$). The magnitude of the non-equilibrium effect factor $|\widetilde{\Lambda}_{\rm{D}}|$ depends on the time averaging window $T$.}
    \label{fig:nonequil_factor}
\end{figure}

Realizations of the non-equilibrium factor $\widetilde{\Lambda}_{\rm{D}} = \langle {(\tilde{\bf{u}} \cdot \nabla) \overline{{\bf{u}}'{}^2}} \rangle$ in the stellar convection simulations are plotted in Figure~\ref{fig:nonequil_factor}. In this case, statistically, the plume velocity is in the downward direction ($\tilde{u}^z < 0$) while the turbulent energy $\overline{{\bf{u}}'{}^2}$ decreases in the downward direction ($\partial \overline{{\bf{u}}'{}^2}/\partial z > 0$). As this result, the non-equilibrium factor $\widetilde{\Lambda}_{\rm{D}}$ is statistically negative as
\begin{equation}
	\widetilde{\Lambda}_{\rm{D}}
	= \left\langle {
		(\tilde{\bf{u}} \cdot \nabla) \overline{{\bf{u}}'{}^2}
	} \right\rangle
	= \left\langle {\tilde{u}}^z 
		\frac{\partial \overline{{\bf{u}}'{}^2}}{\partial z} \right\rangle
	< 0.
	\label{eq:neg_no-equib_factor}
\end{equation}
	
The spatial distribution of $\widetilde{\Lambda}_{\rm{D}}$ is concentrated in the region near surface where the diving plume structures are prominently observed. The magnitude of $\widetilde{\Lambda}_{\rm{D}}$ shows much larger value in the case with smaller averaging time window $T$ defined in (\ref{eq:time_ave_def}). This clearly shows that the non-equilibrium effect associated with the plume motion can be captured by the Lagrangian derivative of the turbulent energy along the plume motion.

In the framework of the time--space averaging procedure, the non-equilibrium effect is incorporated into the turbulent internal-energy flux. On the basis of the eddy-viscosity expression (\ref{eq:nuTN_both_cases}) in the usual ensemble averaging procedure, the turbulent internal-energy diffusivity $\kappa_{\rm{T}}$ with the non-equilibrium effect in the time--space averaging procedure may be formulated as
\begin{equation}
    \kappa_{\textrm{T}}
    = \kappa_{\textrm{TE}} \left( {
        1 - \widetilde{C} \langle {\overline{\rho}} \rangle^{-1/3} 
            \widetilde{\Lambda}_{\rm{D}}
    } \right),
    \label{eq:kappaTN}
\end{equation}
where $\kappa_{\rm{TE}}$ is the equilibrium turbulent internal-energy diffusivity, $\langle {\overline{\rho}} \rangle$ the horizontally averaged density, and $\widetilde{C}$ the model constant. The mean density dependence $\langle {\overline{\rho}} \rangle^{-1/3}$ in (\ref{eq:kappaTN}) is obtained from the argument that the coherent dissipation rate $\widetilde{\varepsilon}$ is proportional to the cubic root of the buoyancy flux associated with a plume. See Section~6.3 and Appendix~C of \citet{yok2022} for the detailed arguments on this dependence.

\begin{figure}[h]
\centering
	\includegraphics[scale=0.7]{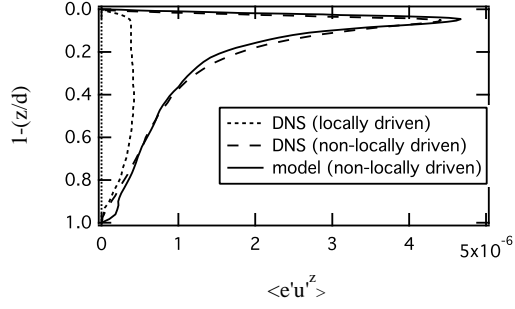}
    \caption{
    Spatial distributions of the turbulent internal-energy flux $\langle {e' u'{}^z} \rangle$ obtained by the direct numerical simulations (DNSs) and the model with the non-equilibrium effect. The left axis $1- (z/d)$ is the depth from the surface ($z=d$). The DNS results for the locally-driven convection case ($\cdots$), for the non-locally-driven convection case ($--$), and the result from the model with the non-equilibrium effect for the non-locally-driven convection case (-----).
    }
    \label{fig:turb_int_en_flux}
\end{figure}

The turbulent internal-energy flux $\langle {e' u'{}^z} \rangle$ obtained from the DNSs in the locally- and non-locally-driven convection cases, which respectively correspond to (a) and (b) of Figure~\ref{fig:entropy_contour_dns}, are plotted in Figure~\ref{fig:turb_int_en_flux}. We see from the figure that the spatial profiles of these two cases are fairly different. First, the magnitude of $\langle {e' u'{}^z} \rangle$ is much larger in the non-locally-driven convection case. At the same time, turbulent transport is much localised near the surface region in the non-locally driven case. These prominent properties of the turbulent transport in the non-locally-driven convection cannot be properly reproduced by the usual gradient diffusion model with the mixing-length theory.

The spatial distribution of $\langle {e' u'{}^z} \rangle$ calculated by the model with the non-equilibrium effect (\ref{eq:kappaTN}):
\begin{equation}
    \langle {e' u'{}^z} \rangle
    = - \kappa_{\rm{T}} \nabla E
    \label{eq:turb_int_en_flux_model}
\end{equation}
is also plotted in Figure~\ref{fig:turb_int_en_flux} ($E$: mean internal energy). The comparison with the DNS results shows that the spatial distribution of $\langle {e' u'{}^z} \rangle$ in the non-locally-driven convection is well reproduced by the present model with the non-equilibrium effect.

\subsection{Interaction between coherent and incoherent fluctuations\label{sec:en_transf}}
In this double-averaging formulation, the fluctuation around the space or ensemble average, $f'$, is divided into the coherent and incoherent components as $f' = \widetilde{f} + f''$. These two components are not independent but interact with each other. For instance, let us consider the Reynolds stress in the time--space double averaging procedure. The evolutions of the Reynolds stress of the coherent velocity fluctuation, $\langle {\overline{\widetilde{\bf{u}} \widetilde{\bf{u}}}} \rangle$, are subject to
\begin{equation}
	\left( {
		\frac{\partial}{\partial t}
		+ \langle {\overline{\bf{u}}} \rangle \cdot \nabla
	} \right) \langle {
		\overline{ {\widetilde{u}}^i {\widetilde{u}}^j}
	} \rangle
	= \widetilde{P}^{ij}
	+ \widetilde{\Pi}^{ij}
	- \widetilde{\varepsilon}^{ij}
	+ \frac{\partial {\widetilde{T}}^{ij\ell}}{\partial x^\ell}
	+ \widetilde{{\cal{P}}}^{ij},
	\label{eq:coherent_rey_strss_eq}
\end{equation}
where $\widetilde{\bf{P}} (= \{ {\widetilde{P}^{ij}} \})$, $\widetilde{\mbox{\boldmath$\Pi$}} (= \{ {\widetilde{\Pi}^{ij}} \})$, $\widetilde{\mbox{\boldmath$\varepsilon$}} (= \{ {\widetilde{\varepsilon}^{ij}} \})$, and $\nabla \cdot \widetilde{\bf{T}} (= \{ {\partial \widetilde{T}^{ij\ell}/\partial x^\ell} \})$ are the production, re-distribution, dissipation, and the transport rates of the coherent components, respectively. They are defined as
\begin{subequations}\label{eq:tilde_rey_budget_terms}
\begin{equation}
	\widetilde{P}^{ij}
	= - \langle {\tilde{u}^j \tilde{u}^\ell} \rangle 
		\frac{\partial \langle {u} \rangle^i}{\partial x^\ell}
	- \langle {\tilde{u}^i \tilde{u}^\ell} \rangle 
		\frac{\partial \langle {u} \rangle^j}{\partial x^\ell},
	\label{eq:tilde_P_ij_def}
\end{equation}
\begin{equation}
	\widetilde{\Pi}^{ij}
	= - \frac{1}{\rho_0} \left\langle {
		\tilde{p} \left( {
			\frac{\partial \tilde{u}^i}{\partial x^j}
			+ \frac{\partial \tilde{u}^j}{\partial x^i}
		} \right)
	} \right\rangle,
	\label{eq:tilde_Pi_ij_def}
\end{equation}
\begin{equation}
	\widetilde{\varepsilon}^{ij}
	- 2 \nu \left\langle {
		\frac{\partial \tilde{u}^i}{\partial x^\ell} 
		\frac{\partial \tilde{u}^j}{\partial x^\ell}
	} \right\rangle,
	\label{eq:tilde_eps_ij_def}
\end{equation}
\begin{equation}
	\frac{\partial \widetilde{T}^{ij\ell}}{\partial x^\ell}
	= + \frac{\partial}{\partial x^\ell} 
		\left( {
			- \left\langle {
				\tilde{u}^\ell \tilde{u}^i \tilde{u}^j
			} \right\rangle
		+ \langle {\tilde{p} \tilde{u}^i} \rangle \delta^{\ell j}
    	+ \langle {\tilde{p} \tilde{u}^j} \rangle \delta^{\ell i}
    	\rule{0.ex}{3.0ex}
  		+ \nu \frac{\partial}{\partial x^\ell}
			\langle {\tilde{u}^i \tilde{u}^j} \rangle
		} \right).
	\label{eq:tilde_T_ij_def}
\end{equation}
\end{subequations}

	On the other hand, the evolution of the Reynolds stress of the incoherent velocity fluctuation, $\langle {\overline{{\bf{u}}'' {\bf{u}}''}} \rangle$, is subject to
\begin{equation}
	\left( {
		\frac{\partial}{\partial t}
		+ \langle {\overline{\bf{u}}} \rangle \cdot \nabla
  		} \right) \langle {
		\overline{ u''{}^i u''{}^j}
	} \rangle
	= P''{}^{ij} 
	+ \Pi''{}^{ij}
	- \varepsilon''{}^{ij}
	+ \frac{\partial T''{}^{ij\ell}}{\partial x^\ell}
	+ {\cal{P}}''{}^{ij},
	\label{eq:_incoh_rey_strss_eq}
\end{equation}
where ${\bf{P}}'' (= \{ {P''{}^{ij}} \})$, $\mbox{\boldmath$\Pi$}'' (= \{ {{\Pi''}^{ij}} \})$, $\mbox{\boldmath$\varepsilon$}'' (= \{ {{\varepsilon''}^{ij}} \})$, and $\nabla \cdot {\bf{T}}'' (= \{ {\partial {T''}^{ij\ell} / \partial x^\ell} \})$ are defined as
\begin{subequations}\label{eq:rey''_budget_terms}
\begin{equation}
	P''{}^{ij}
	= - \langle {u''{}^j u''{}^\ell} \rangle 
		\frac{\partial \langle {u} \rangle^i}{\partial x^\ell}
	- \langle {u''{}^i u''{}^\ell} \rangle 
		\frac{\partial \langle {u} \rangle^j}{\partial x^\ell},
	\label{eq:P''_ij_def}
\end{equation}
\begin{equation}
	\Pi''{}^{ij}
	= - \frac{1}{\rho_0} \left\langle {
		p'' \left( {
			\frac{\partial u''{}^i}{\partial x^j}
			+ \frac{\partial u''{}^j}{\partial x^i}
		} \right)
	} \right\rangle
	- 2 \nu \left\langle {
		\frac{\partial u''{}^i}{\partial x^\ell} 
		\frac{\partial u''{}^j}{\partial x^\ell}
	} \right\rangle,
	\label{eq:Pi''_ij_def}
\end{equation}
\begin{equation}
	\varepsilon''{}^{ij}
	= - 2 \nu \left\langle {
		\frac{\partial u''{}^i}{\partial x^\ell} 
		\frac{\partial u''{}^j}{\partial x^\ell}
	} \right\rangle,
	\label{eq:eps''_ij_def}
\end{equation}
\begin{equation}
	\frac{\partial T''{}^{ij\ell}}{\partial x^\ell}
	= + \frac{\partial}{\partial x^\ell} 
		\left( {
	- \left\langle {
		u''{}^\ell u''{}^i u''{}^j
	} \right\rangle
    + \langle {p'' u''{}^i} \rangle \delta^{\ell j}
	+ \langle {p'' u''{}^j} \rangle \delta^{\ell i}
	+ \nu \frac{\partial}{\partial x^\ell}
		\langle {u''{}^i u''{}^j} \rangle
	} \right).
	\label{eq:T''_ij_def}
\end{equation}
\end{subequations}
The terms (\ref{eq:tilde_rey_budget_terms}) and (\ref{eq:rey''_budget_terms}) are similar to the counterparts of the Reynolds stress equation without using the double averaging procedure. Actually, for example, the addition of $\widetilde{P}^{ij}$ and $P''{}^{ij}$ just gives the usual production term of $\langle {u'{}^i u'{}^j} \rangle$. On the other hand, the final terms in (\ref{eq:coherent_rey_strss_eq}) and (\ref{eq:_incoh_rey_strss_eq}) are originated from the double-averaging procedure. These terms, ${\widetilde{\cal{P}}}^{ij}$ and ${\cal{P}}''{}^{ij}$, are defined by
\begin{equation}
	{\cal{P}}''{}^{ij}
	= - \left\langle {
    	\widetilde{u''{}^\ell u''{}^j} 
    	\frac{\partial \widetilde{u}^i}{\partial x^\ell}
  	} \right\rangle
	- \left\langle {
    	\widetilde{u''{}^\ell u''{}^i} 
    	\frac{\partial \widetilde{u}^j}{\partial x^\ell}
	} \right\rangle
	= - \widetilde{\cal{P}}^{ij},
	\label{eq:coh_incoh_int_act}
\end{equation}
and represent the transfer rates between the coherent and incoherent components. Here, the dispersion part of the incoherent Reynolds stress is defined by
\begin{equation}
	\widetilde{u''{}^i u''{}^j}
	= \overline{u''{}^i u''{}^j} 
	- \left\langle {\overline{u''{}^i u''{}^j}} \right\rangle.
	\label{eq:dispers_rey_strss_def}
\end{equation}
If we add (\ref{eq:coherent_rey_strss_eq}) and (\ref{eq:_incoh_rey_strss_eq}), these two terms cancels with each other, and will not contribute to the budget of the total Reynolds stress $\langle {{\bf{u}}' {\bf{u}}'} \rangle$ but to the transfer between the coherent and incoherent components. The dispersion part of the incoherent Reynolds stress $\widetilde{u''{}^\ell u''{}^j}$ (\ref{eq:dispers_rey_strss_def}) coupled with the spatial structure of the coherent motion, $\partial \widetilde{u}^j / \partial x^\ell$, the turbulent stress are transferred between the coherent and incoherent components. 

\begin{figure}
\centering
	\includegraphics[scale=0.7]{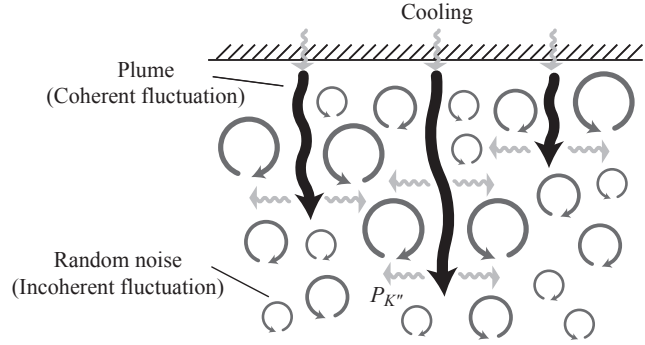}
    \caption{Schematic picture of the interaction between the coherent and incoherent fluctuations. Plumes (coherent fluctuations, depicted by thick balack curved lines) are driven by the surface cooling. The energy of the plume motions are transferred to the energy of the random noise (incoherent fluctuation, depicted by grey circle eddies) by the interaction $P_{K''}$ if $P_{K''} > 0$.}
    \label{fig:coh_incoh_int_act}
\end{figure}

Taking the contraction of $i$ and $j$ in (\ref{eq:coherent_rey_strss_eq}) and (\ref{eq:_incoh_rey_strss_eq}), we obtain the evolution equations of the coherent and incoherent energies as
\begin{equation}
	\left( {
		\frac{\partial}{\partial t}
		+ \langle {\overline{\bf{u}}} \rangle \cdot \nabla
	} \right) \left\langle {
		\frac{1}{2} {\tilde{\bf{u}}}^2
	} \right\rangle
	= \widetilde{P}
	- \widetilde{\varepsilon}
	+ \nabla \cdot \widetilde{\bf{T}}
	+ \left\langle {
		\widetilde{u''{}^\ell u''{}^m} 
		\frac{\partial \widetilde{u}^m}{\partial x^\ell}
	} \right\rangle,
	\label{eq:coh_en_eq}
\end{equation}
\begin{equation}
	\left( {
		\frac{\partial}{\partial t}
		+ \langle {\overline{\bf{u}}} \rangle \cdot \nabla
	} \right) \left\langle {
		\frac{1}{2}{\bf{u}}''{}^2
	} \right\rangle
	= P''
	- \varepsilon''
	+ \nabla \cdot {\bf{T}}'' 
	- \left\langle {
		\widetilde{u''{}^\ell u''{}^m} 
		\frac{\partial \widetilde{u}^m}{\partial x^\ell}
	} \right\rangle,
	\label{eq:incoh_en_eq}
\end{equation}
where $\widetilde{P}$ and $P''$ are the production rates, $\widetilde{\varepsilon}$ and $\varepsilon''$ the dissipation rates, $\nabla \cdot \widetilde{\bf{T}}$ and $\nabla \cdot {\bf{T}}''$ the transport rates of the coherent and incoherent turbulent energies, respectively. The final terms in (\ref{eq:coh_en_eq}) and (\ref{eq:incoh_en_eq}) represent the energy transfer rates between the coherent and incoherent turbulent energies. They are denoted as
\begin{equation}
	{P}_{K''}
	= - \left\langle {
		\widetilde{u''{}^\ell u''{}^m} 
		\frac{\partial \widetilde{u}^m}{\partial x^\ell}
	} \right\rangle
	= - {P}_{\tilde{K}}.
	\label{eq:P_K''_P_tildeK}
\end{equation}
If we approximate the dispersion part of the Reynolds stress by the eddy-viscosity-type model by
\begin{equation}
	\widetilde{u''{}^\ell u''{}^m} 
	\simeq - \widetilde{\nu} 
		\frac{\partial \widetilde{u}^m}{\partial x^\ell},
	\label{eq:eddy_visc_model_dis_rey}
\end{equation}
the energy transfer rate from the coherent to incoherent fluctuations, $P_{K''}$, is expressed as
\begin{equation}
	{P}_{K''} 
	= - \left\langle {
        \widetilde{u''{}^\ell u''{}^m} 
		\frac{\partial \widetilde{u}^m}{\partial x^\ell}
    } \right\rangle
	\simeq + \widetilde{\nu} \left( {
		\frac{\partial \widetilde{u}^m}{\partial x^\ell}
	} \right)^2
	> 0.
	\label{eq:pos_PK''}
\end{equation} 
As (\ref{eq:P_K''_P_tildeK}) shows, the sink of the coherent fluctuation energy, ${P}_{\tilde{K}} <0$, corresponds to the production of the incoherent fluctuation energy, ${P}_{K''} >0$. In this case, the kinetic energy of the plume or coherent fluctuation motion driven by the surface cooling is transferred towards the kinetic energy of the random or incoherent fluctuation motions as schematically depicted in Figure~\ref{fig:coh_incoh_int_act}.

In this picture, the diving plumes driven by the surface cooling enhances the turbulent transport through the non-equilibrium effect along the plume motions. During this process, the energy of plumes (coherent fluctuations) is transferred to the energy of random noises (incoherent fluctuations) through the inhomogeneous motions of plumes $\nabla  \widetilde{\bf{u}}$ coupled with the incoherent fluctuations $\widetilde{{\bf{u}}'' {\bf{u}}''}$. Of course, there are several other mechanisms such as the production $P''$, dissipation $\varepsilon''$, etc.\ in (\ref{eq:incoh_en_eq}), that contribute to the evolution of the incoherent fluctuation energy $\langle {{\bf{u}}''{}^2} \rangle$. In this sense, the present picture is a fairly simplified one. 

    With the nonequilibrium effect, the energy transfer rate to the incoherent component of the fluctuation is written as
\begin{equation}
	{P}_{K''} 
	= - \left\langle{ 
        \widetilde{u''{}^\ell u''{}^m} 
		\frac{\partial \widetilde{u}^m}{\partial x^\ell}
    } \right\rangle
	\simeq 
    + \widetilde{\nu} \left( {1 - \widetilde{\Lambda}_{\rm{D}}} \right)
    \left( {
		\frac{\partial \widetilde{u}^m}{\partial x^\ell}
	} \right)^2.
	\label{eq:pos_PK''_NE}
\end{equation}
As the DNSs of the surface cooling case show (Figure~\ref{fig:nonequil_factor}), the non-equilibrium-effect coefficient $\widetilde{\Lambda}_{\rm{D}}$ is negative:
\begin{equation}
    \widetilde{\Lambda}_{\rm{D}} < 0.
    \label{eq:Lambda_neg}
\end{equation}
In this case, the energy transfer to the incoherent or random fluctuation is enhanced by the non-equilibrium turbulence effect along the plume motion. This matches the tendency of the surface cooling-driven convection case. The presence of the diving plumes strongly enhances the turbulence mixing through the non-equilibrium effect. This picture is based on several simplification, but still suggests the importance of the non-equilibrium effect associated with the plume motions in the turbulent transport in convection.

\section{Conclusion \label{sec:concl}}
Linear gradient-transport-type turbulence models for the turbulent fluxes with the eddy-viscosity and -diffusivity representation work poorly in some turbulent flows. For example, the $K-\varepsilon$ turbulence model based on the eddy-viscosity $\nu_{\textrm{T}} = C_\nu K^2/\varepsilon$ cannot be applied to the homogeneous-shear turbulence; it leads to the overestimation of $K$ and $\varepsilon$. Convective turbulence with plumes/jets and supersonic turbulence accompanying shocks constitute other examples. Simulations with a simple linear model provide substantially over- or under-estimate of the turbulent transports (turbulent viscosity, mass flux, heat flux, etc.). To rectify this deficiency of linear gradient-transport models, in this work, the non-equilibrium effect mediated by streamwise variation of turbulence are incorporated into the turbulence model. A turbulence model with these effects implemented in combination was proposed. In the model, in addition to the usual eddy-viscosity and -diffusivity representations for the turbulent transports, deviations from the simple representations due to the non-equilibrium effect is taken into account. 

Relevance of the non-equilibrium  effects was argued with utilising recent elaborated experiments in turbulent jets and plumes. It was pointed out that the expected results of the non-equilibrium effect on turbulent transports in the turbulent jets and plumes are transport enhancement or suppression due to the non-equilibrium effect (depending on the turbulence variation along the jets/plumes). These evaluations of turbulent transports seem to be in preferable directions to correct the discrepant results obtained from simple conventional gradient-transport-type models.

\section*{Funding}
{This research was funded by Japan Society for the Promotion of Science (JSPS) Grants-in-Aid for Scientific Research: JP18H01212 and JP23H01199. 
The Issac Newton Institute (INI) for Mathematical Sciences program: ``Frontier in dynamo theory: from the Earth to the stars'' (DYT2) was supported by EPSRC grant no EP/R014604/1. The Nordic Institute for Theoretical Physics (NORDITA) program: ``Toward a comprehensive model for the galactic magnetic field'' was partly supported by NordForsk and Royal Astronomical Society.
}

\acknowledgments{The author would like to thank Youhei Masada and Tomoya Takiwaki for 
providing data of numerical simulations and
valuable discussions on density-variance and non-equilibrium effects in stellar convection simulations. 
Part of this work was started during the author's stay at the INI, Cambridge, during the program: DYT2 from September to December 2022. The author also would like to thank the NORDITA for support and hospitality during the program: Towards a comprehensive model for the galactic magnetic field, in April 2023.
}

\section*{Data Availability}
Data of the numerical simulations and analysis presented in this paper are available from the author upon reasonable request.

\appendix

\section{Non-equilibrium turbulent energy from the two-scale direct-interaction approximation analysis\label{sec:append_A}}
From the two-scale direct-interaction approximation (TSDIA) analysis, the turbulent energy $K (\equiv \langle {{\textbf{u}}'{}^2} \rangle/2)$ is known to be expressed in terms of as \citep{yos1984}
\begin{equation}
    K = \int d{\textbf{k}}\ Q(k; \tau,\tau)
    - \int d{\textbf{k}} \int_{-\infty}^{\tau}\!\! d\tau_1\ 
        G(k,{\textbf{X}};\tau,\tau_1)
        \frac{DQ(k,{\textbf{X}};\tau,\tau_1)}{DT},
    \label{eq:K_exp_tsdia}
\end{equation}
where $Q(k, {\textbf{X}}; \tau,\tau',T)$ and $G(k, {\textbf{X}}; \tau,\tau',T)$ are the energy spectral function and the response function defined by
\begin{equation}
    Q(k,{\textbf{X}};\tau,\tau',T)
    = \sigma(k,{\textbf{X}};T)
        \exp \left[ {- \omega(k,{\textbf{X}};T) |\tau - \tau'|} \right],
    \label{eq:Q_exp}
\end{equation}
\begin{equation}
    G(k,{\textbf{X}};\tau,\tau',T)
    = H(k,\tau-\tau')
        \exp \left[ {
            - \omega(k,{\textbf{X}};T) (\tau - \tau')
        } \right],
    \label{eq:G_exp}
\end{equation}
respectively. Here the Heaviside step function defined by
\begin{equation}
    H(x) 
    = \left\{ {
    \begin{array}{ll}
        1\;\; &\mbox{for}\;\; x \ge 0,\\
        0\;\; &\mbox{for}\;\; x < 0,
    \end{array}
    } \right.
    \label{eq:heaviside_fn_def}
\end{equation}
and
\begin{equation}
    \sigma(k,{\textbf{X}};T)
    = \sigma_0 \varepsilon({\textbf{X}},T)^{2/3} k^{-11/3},
    \label{eq:sigma_scaling}
\end{equation}
\begin{equation}
    \omega(k,{\textbf{X}};T)
    = \omega_0 \varepsilon({\textbf{X}},T)^{1/3} k^{2/3}.
    \label{eq:omega_scaling}
\end{equation}
As we see below, eq.~(\ref{eq:sigma_scaling}) corresponds to the Kolmogorov scaling of homogeneous and isotropic equilibrium turbulence.

Substituting (\ref{eq:Q_exp})-(\ref{eq:omega_scaling}) into eq.~(\ref{eq:K_exp_tsdia}), we have
\begin{eqnarray}
    K
    &=& \int\! d{\textbf{k}}\
        \sigma({\textbf{k}},{\textbf{X}};T)
        \exp[- \omega(k,{\textbf{X}};T) |\tau - \tau|]
    \nonumber\\
    &-& \int\! d{\textbf{k}} \int_{-\infty}^\tau\!\!\! d\tau_1 
        H(\tau - \tau_1) \exp[-\omega(k,{\textbf{X}};T)(\tau - \tau_1)]
    \nonumber\\
    &&\hspace{20pt} \times \frac{D}{DT} \sigma(k,{\textbf{X}};T)
        \exp[-\omega(k,{\textbf{X}};T)|\tau - \tau_1|]
    \label{eq:K_exp_sigma_omega}
\end{eqnarray}

The first term of (\ref{eq:K_exp_sigma_omega}) denoted by $K_{\textrm{E}}$ is evaluated as
\begin{eqnarray}
    K_{\textrm{E}}
    &=& \int\! d{\textbf{k}}\
        \sigma({\textbf{k}},{\textbf{X}};T)
        \exp[- \omega(k,{\textbf{X}};T) |\tau - \tau|]
    \nonumber\\
    &=& 4\pi \int_{k_{\textrm{C}}}^{\infty}\!\! dk\ k^2
        \sigma_0 \varepsilon^{2/3} k^{-11/3}
    = 4\pi \sigma_0 \int_{k_{\textrm{C}}}^{\infty}\!\! dk\
        \varepsilon^{2/3} k^{-5/3},
    \label{eq:K_E_exp_sigma_omega}
\end{eqnarray}
where
\begin{equation}
    k_{\textrm{C}} = 2\pi / \ell_{\textrm{C}}
    \label{eq:k_cutoff_def}
\end{equation}
is the infrared cut-off wave number representing the largest eddy size of turbulence, $\ell_{\textrm{C}}$. We further introduce scale transformation
\begin{equation}
    s = k / k_{\textrm{C}}
    \label{eq:s_k_over_kc}
\end{equation}
to express the relative scale based on the infrared cut-off wave number. With this length scale, the equilibrium energy part can be expressed as
\begin{eqnarray}
    K_{\textrm{E}}
    &=& 4 \pi \sigma_0 \varepsilon^{2/3} k_{\textrm{C}} 
        \int_{s \ge 1}\!\! ds\ s^{-5/3}
    = 6 \pi \sigma_0 \varepsilon^{2/3} k_{\textrm{C}}^{-2/3}
    \nonumber\\
    &=& 3 (2\pi)^{1/3} \sigma_0 \varepsilon^{2/3}
        \ell_{\textrm{C}}^{2/3}.
    \label{KE_epsilon_ell}
\end{eqnarray}
The second term in (\ref{eq:K_exp_sigma_omega}) represents the non-equilibrium effect on the turbulent energy. This part is evaluated as
\begin{eqnarray}
    K_{\textrm{N}}
    &=& - \int\! d{\textbf{k}} \int_{-\infty}^\tau\!\!\! d\tau_1 
        H(\tau - \tau_1) \exp[-\omega(k,{\textbf{X}};T)(\tau - \tau_1)]
    \nonumber\\
    &&\hspace{20pt} \times \frac{D}{DT} \sigma(k,{\textbf{X}};T)
        \exp[-\omega(k,{\textbf{X}};T)|\tau - \tau_1|]
    \nonumber\\
    &=& - \int\! d{\textbf{k}} \int_{-\infty}^\tau\!\!\! d\tau_1
        \frac{D\sigma(k,{\textbf{X}};T)}{DT}
        \exp[-2\omega(k,{\textbf{X}};T)(\tau - \tau_1)] 
    \nonumber\\
    &&\hspace{10pt}+ \int\! d{\textbf{k}} \int_{-\infty}^\tau\!\!\!         d\tau_1 (\tau - \tau_1) \sigma(k,{\textbf{X}};T)
            \frac{D\omega}{DT} 
            \exp[-2\omega(k,{\textbf{X}};T)(\tau - \tau_1)]
    \nonumber\\
    &=& - \left[ {
        4\pi \int k^2 dk \frac{1}{2\omega} \frac{D\sigma}{DT}
        \exp[-2\omega (k,{\textbf{X}},T)(\tau-\tau_1)]
    }\right]_{\tau_1 = - \infty}^\tau
    \nonumber\\
    &&\hspace{10pt} + \left[ {
        4\pi \int dk\ k^2 \frac{\sigma}{2 \omega^2} \frac{D\omega}{DT} \exp[-2\omega(k,{\textbf{X}};T)(\tau - \tau_1)]
    } \right]_{\tau_1 = -\infty}^\tau
    \nonumber\\
    &=& -2\pi \int dk\ k^2 \left( {
        \frac{1}{\omega} \frac{D\sigma}{DT}
        - \frac{\sigma}{2\omega^2} \frac{D\omega}{DT}
    } \right)
    \label{eq:KN_sigma_omega}
\end{eqnarray}
Here, in integration with respect to $\tau_1$, we have integrated by parts, and used the fact that neither $\sigma$ nor $\omega$ explicitly depends on $\tau_1$. With eqs.~(\ref{eq:sigma_scaling}) and (\ref{eq:omega_scaling}), (\ref{eq:KN_sigma_omega}) is calculated as
\begin{equation}
    K_{\textrm{N}}
    = - \pi \frac{\sigma_0}{\omega_0} k_{\textrm{C}}^{-4/3}
        \varepsilon^{-2/3} \frac{D\varepsilon}{DT}
    + \frac{\pi}{6} \frac{\sigma_0}{\omega_0} 
        \varepsilon^{1/3} k_{\textrm{C}}^{-7/3}
        \frac{Dk_{\textrm{C}}}{DT}.
    \label{eq:KN_exp_kc_eps}
\end{equation}	
It follows from eq.~(\ref{eq:k_cutoff_def}) that the non-equilibrium part of the energy is expressed in terms of $\varepsilon$ and $\ell_{\textrm{C}}$ as
\begin{equation}
    K_{\textrm{N}}
    = - \frac{1}{2} (2\pi)^{-1/3} \frac{\sigma_0}{\omega_0}  
        \varepsilon^{-2/3} \ell_{\textrm{C}}^{4/3} 
        \frac{D\varepsilon}{DT}
    - \frac{1}{12} (2\pi)^{-4/3} \frac{\sigma_0}{\omega_0}
        \varepsilon^{1/3} \ell_{\textrm{C}}^{1/3}
        \frac{D\ell_{\textrm{C}}}{DT}.
    \label{eq:KN_exp_ell_eps}
\end{equation}
Added with (\ref{KE_epsilon_ell}), the turbulent energy is expressed as
\begin{equation}
    K
    = 3 (2\pi)^{1/3} \sigma_0 
        \varepsilon^{2/3} \ell_{\textrm{C}}^{2/3}
    - \frac{1}{2} (2\pi)^{-1/3} \frac{\sigma_0}{\omega_0}  
        \varepsilon^{-2/3} \ell_{\textrm{C}}^{4/3} 
        \frac{D\varepsilon}{DT}
    - \frac{1}{12} (2\pi)^{-4/3} \frac{\sigma_0}{\omega_0}
        \varepsilon^{1/3} \ell_{\textrm{C}}^{1/3}
        \frac{D\ell_{\textrm{C}}}{DT},
    \label{eq:K_ell_eps}
\end{equation}
or with model constants $C_{K1}$, $C_{K2}$, and $C_{K3}$, the expression for the turbulent kinetic energy is written as
\begin{equation}
    K 
    = C_{K1} \varepsilon^{2/3} \ell_{\textrm{C}}^{2/3}
    - C_{K2} \varepsilon^{-3/2} \ell_{\textrm{C}}^{4/3} 
        \frac{D\varepsilon}{Dt}
    - C_{K3} \varepsilon^{1/3} \ell_{\textrm{C}}^{1/3}
        \frac{D\ell_{\textrm{C}}}{Dt},
    \label{eq:K_exp_ell_eps_Cn}
\end{equation}
where the model constants are defined as
\begin{equation}
    C_{K1} = 3 (2\pi)^{1/3} \sigma_0,\;\;
    C_{K2} = \frac{1}{2} (2\pi)^{-1/3} \frac{\sigma_0}{\omega_0},\;\;
    C_{K3} = \frac{1}{12} (2\pi)^{-4/3} \frac{\sigma_0}{\omega_0}.
    \label{eq:C_Kn_constants}
\end{equation}
Equation~(\ref{eq:K_exp_ell_eps_Cn}) is Eq.~(\ref{eq:K_exp_eps_ell}) in Sec.~\ref{sec:noneq_effect}.







\end{document}